\documentstyle[floats,multicol,aps,epsf,prl]{revtex}
\begin{document}
\draft
\twocolumn[\hsize\textwidth\columnwidth\hsize\csname @twocolumnfalse\endcsname
\title{Intrinsic temperature dependences of transport coefficients within the
hot-spot model for normal state YBCO} 
\author{J. Paaske}
\address{{\O}rsted Laboratory, Niels Bohr Institute, University of Copenhagen,\\
 Universitetsparken 5, DK-2100 Copenhagen {\O}, Denmark}
\author{D. V. Khveshchenko}
\address{NORDITA, Blegdamsvej 17, DK-2100 Copenhagen {\O}, Denmark}
\date{\today}
\maketitle
\begin{abstract}
The temperature dependences of the galvanomagnetic and thermoelectric
transport coefficients within a generic {\it hot-spot} model are
reconsidered. Despite the recent success in explaining ac Hall effect data
in YBa$_{2}$Cu$_{3}$O$_{7}$, a general feature of this model is a
departure from the approximately universal temperature dependences observed
for normal state transport in the optimally doped cuprates. In this paper,
we discuss such systematic deviations and illustrate some of their effects
through a concrete numerical example using the calculated band structure for
YBa$_{2}$Cu$_{3}$O$_{7}$.
\end{abstract}
\vskip2pc
]

The construction of a consistent phenomenology for normal state transport in
the cuprates has turned out to be highly problematic. At an early stage it
was suggested by Anderson~\cite{pwa} that experimental magnetotransport data
should be fitted using two distinct relaxation rates, and plentiful experimental
verifications have by now made this conjecture an attractive ansatz for 
interpreting normal state transport data. Thus, in pure optimally doped YBCO,
resistivity and inverse Hall angle have been successfully related to the two
relaxation rates
\begin{equation}
1/\tau_{tr}=\eta T ,\hspace*{1cm} 1/\tau_{H}=T^{2}/W_{H},~\label{eq:relaxt}
\end{equation}
proving to be valid over a substantial range of temperatures~\cite{harris1}.
The prefactor $\eta$ may be inferred from the width of the ac conductivity peak
to be $\sim$2~\cite{orenstein} while the energy scale $W_{H}$ is less agreed 
upon. The original Anderson proposal related $W_{H}$ to the spinon bandwidth,
of the order of the superexchange energy $J \sim 1400$ K, which is in reasonable 
accordance with an experimental fit by Chien {\it et al.}~\cite{chien} who found
$W_{H}$ to be 830 K. This value however presupposes a connection between $W_{H}$
and the cyclotron frequency, and as pointed out in Ref.~\cite{2xlee}, the same
data yield $W_{H}\sim 65$ K or less, if in the cyclotron frequency one uses
simply the mass deduced from the optical conductivity. The latter interpretation
complies well with the ac magnetotransport data of Drew
{\it et al.}~\cite{drew,parks} suggesting that $W_{H}\sim 120$ K.

Various phenomenologies have been suggested as underlying this experimentally 
appealing concept of two relaxation rates, but, as pointed out in a recent work 
by Coleman, Schofield and Tsvelik~\cite{coleman}, most of these suggestions are 
problematic in one way or the other. They pointed to the fact that electrical
current and Hall current have opposite parity under charge-conjugation, which
seems to indicate that some as yet unknown scattering mechanism sensitive to 
charge-conjugation is at play. Their Majorana fermion Boltzmann equation yields 
a set of transport coefficients, in which the scattering rates combine in
{\it series} or {\it parallel}, making either the smallest or the largest of
the two dominant and therefore it relies on a marked difference in magnitude
of the two rates. A particularly interesting consequence of this phenomenology
is the fact that the thermopower turns out to be simply related to the Hall
constant as 
\begin{equation}
\frac{S}{T}=a+b R_{H},
\end{equation}
implying that a longitudinal current, such as the response to a temperature
gradient, may also be related to the Hall scattering rate. This appears to be
consistent with thermoelectric experiments in thin films of
Tl$_2$Ba$_2$CaCu$_2$O$_{8+\delta}$~\cite{clay}, supporting the idea that the
two relaxation rates do not pertain to transverse and longitudinal currents 
respectively, but rather to currents of even or odd charge conjugation parity.

In the course of fitting  ac Hall effect data the standard Bloch-Boltzmann 
theory based on bandstructure calculations and angular resolved photoemission 
experiments, supplied by the assumption of an anisotropic relaxation rate, has
been reconsidered recently by Zheleznyak, Yakovenko, Drew and
Mazin~\cite{zheleznyak}. They have elaborated on what they name the
{\it additive} two-$\tau$ approach taken earlier by Carrington
{\it et al.}~\cite{carrington} and Kendziora {\it et al.}~\cite{kendziora},
where the relaxation rate was assumed to have different temperature dependences
on different parts of the Fermi surface. The notion of an anisotropic
relaxation rate has been applied succesfully in explaining the observed
deviations from nearly free electron values of the Hall constant in Al and
Pd~\cite{schulz}, and seems as a natural minimal model to explain abnormal
transport data. However, as the present Communication is ment to demonstrate,
this approach leads to temperature dependences which are largely inconsistent
with normal state transport experiments in optimally doped cuprates.

\begin{figure}[tb]
\epsfxsize=5.5cm
\epsfysize=5.0cm
\centerline{\epsfbox{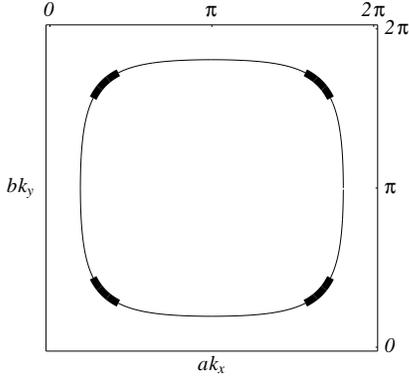}}
\vspace{1cm}
\caption[Figure 1]{\protect Even plane band Fermi surface for
YBa$_{2}$Cu$_{3}$O$_{7}$, from Ref.~\cite{andersen}. In the text, flat parts
(thin lines) and corners are referred to by subscripts $f$ and $c$ respectively.}
\label{fig1}
\end{figure}

The authors of Ref.~\cite{zheleznyak} have mainly focused on frequency 
dependences, and for this purpose indeed the additive two-$\tau$ model works
rather well. Their analysis was based on the band structure calculations for 
YBa$_{2}$Cu$_{3}$O$_{7}$ by Andersen {\it et al}.~\cite{andersen}, supplied
with the assumption that the relaxation rate has linear temperature dependence
on the flat parts and quadratic temperature dependence on the corners of the
Fermi surface. Only the in-plane bonding band was considered since it was
found to have the largest contribution to the conductivities. With the 
parameters given in Ref.~\cite{andersen} for this even plane band, the
resulting Fermi surface has been replotted in Fig.~\ref{fig1} and the notion
of flat parts and corners are seen to be easily resolved.
Continuing along the lines of Ref.~\cite{zheleznyak}, we now want to focus on
temperature rather than frequency dependences.

Within the {\it momentum independent} relaxation time approximation, a standard
solution of the Boltzmann equation for a square symmetric 2D system (neglecting
the slight anisotropy of YBa$_{2}$Cu$_{3}$O$_{7}$) yields the following low
temperature and low field longitudinal, Hall, thermoelectric and transverse
magnetoconductivity:
\begin{eqnarray}
\sigma&=&\left(\frac{e}{2\pi}\right)^{2}\oint\!d k_{\|}
v_{i}\delta_{ij}v_{j}\tau/v,\label{eq:sigm0}\\
\sigma_{H}&=&-eB\left(\frac{e}{2\pi}\right)^{2}\oint\!d k_{\|}
v_{i}\epsilon_{ik}m^{-1}_{kl}\epsilon_{lj}v_{j}\tau^{2}/v,\\
\beta&=&-e{\cal L}_{0}T\left(\frac{e}{2\pi}\right)^{2}\oint\!d k_{\|}
v_{i}m^{-1}_{ij}v_{j}\tau/v^{3},\\
\Delta\sigma&=&-(eB)^{2}\left(\frac{e}{2\pi}\right)^{2}\oint\!d k_{\|}
v_{i}\epsilon_{ik}m^{-1}_{kn}m^{-1}_{nl}\epsilon_{lj}v_{j}\tau^{3}/v,
\label{eq:deltsigm0}
\end{eqnarray}
expressed in terms of the inverse mass tensor
$m_{ij}^{-1}=\frac{\partial^{2}\varepsilon}{\partial_{k_{i}}\partial_{k_{j}}}$, 
the Levi-Civit\`{a} tensor $\epsilon_{ij}$ and the Lorenz number
${\cal L}_{0}=\pi^{2}/3e^{2}$. The Fermi surface has been parametrized by 
arclength $k_{\|}$ and the two dimensional momentum integrals carried out as
$\int\!d^{\,2}\!k\ldots=\int\!d\varepsilon\oint\!\frac{dk_{\|}}{v}\ldots$. The
thermoelectric conductivity rests upon a first order expansion of the velocity 
in local momentum coordinates along the normal and tangential directions of the 
constant energy contours
\begin{equation}
v=v_{F}+\left(\frac{v_{i}}{v}\frac{\partial v_{i}}{\partial k_{j}}\frac{v_{j}}
{v}\right)\delta k_{\bot}+\left(\frac{\epsilon_{ij}v_{j}}{v}\frac{\partial 
v_{l}}{\partial k_{i}}\frac{v_{l}}{v}\right)\delta k_{\|},
\end{equation}
where only the normal component proportional to $\delta \varepsilon=v\delta 
k_{\bot}$ provides a nonzero contribution to the thermoelectric conductivity.

In the case of an anisotropic relaxation time $\tau({\bf k})$, the first three 
transport coefficients listed above are only modified by the fact that $\tau$
now depends on $k_{\|}$. Using a somewhat oversimplified momentum dependence
of $\tau$, the hot-spot Fermi surface is implemented via step functions for
the flat parts and corners multiplied by $\tau_{f}=1/\eta T$ and
$\tau_{c}=W_{H}/T^{2}$  respectively. These particular dependences are largely
motivated by both the expected outcome in terms of the transport relaxation
rates from Eq.($\!\!$~\ref{eq:relaxt}) and the available (semi) microscopic
calculations for Fermi surfaces with nearly flat parts, van Hove singularities, 
or in the presence of commensurate antiferromagnetic fluctuations. Introducing
weight factors of the flat parts (see Fig.~\ref{fig1})
\begin{eqnarray}
a&=&\frac{\oint_{f}\!d k_{\|} v}{\oint\!d k_{\|} v},\\
b&=&\frac{\oint_{f}\!d k_{\|}
\left(v_{i}\epsilon_{ik}m^{-1}_{kl}\epsilon_{lj}v_{j}\right)/v}{\oint\!d k_{\|}
\left(v_{i}\epsilon_{ik}m^{-1}_{kl}\epsilon_{lj}v_{j}\right)/v},\\
c&=&\frac{\oint_{f}\!d k_{\|}v_{i}m^{-1}_{ij}v_{j}/v^{3}}{\oint\!d k_{\|}
v_{i}m^{-1}_{ij}v_{j}/v^{3}},
\end{eqnarray}
the plasma frequency
\begin{equation}
\omega_{p}^{2}=\left(\frac{e}{2\pi}\right)^{2}\oint\!d k_{\|} v,
\end{equation}
a generalized cyclotron frequency
\begin{equation}
\omega_{H}=\frac{-eB\oint\!d k_{\|}\left(v_{i}\epsilon_{ik}m^{-1}_{kl}
\epsilon_{lj}v_{j}\right)/v}{\oint\!d k_{\|} v}
\end{equation}
and the energy scale 
\begin{equation}
\omega_{0}=\frac{\oint\!d k_{\|} v}
           {\oint\!d k_{\|}v_{i}m^{-1}_{ij}v_{j}/v^{3}},
\end{equation}
which for an isotropic system reduces to the Fermi energy, the longitudinal, 
Hall and thermoelectric conductivities may be written as
\begin{eqnarray}
\sigma&=&\omega_{p}^{2}\left[a\tau_{f}+(1-a)\tau_{c}\right],\label{eq:cond1}\\
\sigma_{H}&=&\omega_{p}^{2}\omega_{H}\left[b\tau_{f}^{2}+(1-b)\tau_{c}^{2}\right],
\label{eq:hcond}\\
\beta&=&-(e{\cal L}_{0}T\omega_{p}^{2}/\omega_{0})[c\tau_{f}+(1-c)\tau_{c}].
\label{eq:cond3}
\end{eqnarray}

It is the square symmetry which makes the $\partial_{k\|}\tau$ terms cancel in
$\sigma_{H}$ and with the expansion used in $\beta$ no such terms appear in the 
first place. As for the magnetoconductivity, however, the integrand will in
general acquire additional terms proportional to powers of $\partial_{k\|}\tau$,
the contributions of which are beyond the underlying step function approximation
for $\tau({\bf k})$.

With temperature scales $T^{\ast}_{a}=(\frac{1-a}{a})\eta W_{H}$,
$T^{\ast}_{b}=(\frac{1-b}{b})^{1/2}\eta W_{H}$ and
$T^{\ast}_{c}=(\frac{1-c}{c})\eta W_{H}$
Eqs.($\!\!$~\ref{eq:cond1}-$\!\!$~\ref{eq:cond3}) lead to the following 
temperature dependences given by simple fractions
\begin{eqnarray}
\rho&=&\frac{\eta}{\omega_{p}^{2}a}\frac{T}{1+T^{\ast}_{a}/T},\label{eq:cond4}\\
\sigma_{H}&=&\frac{\omega_{p}^{2}\omega_{H}b}{\eta^{2}}
\frac{1+(T^{\ast}_{b}/T)^{2}}{T^{2}},\label{eq:hcond2}\\
\cot\theta_{H}&=&\frac{\eta a}{\omega_{H}b}\frac{1+T^{\ast}_{a}/T}
{1+(T^{\ast}_{b}/T)^{2}}T,\\
R_{H}&=&\frac{\omega_{H}b}{B\omega_{p}^{2}a^{2}}\frac{1+(T^{\ast}_{b}/T)^{2}}
{\left(1+T^{\ast}_{a}/T\right)^{2}},\\
\beta&=&-\frac{e{\cal L}_{0}\omega_{p}^{2}c}{\eta\omega_{0}}
\left(1+T^{\ast}_{c}/T\right),\\
S&=&\frac{e{\cal L}_{0}c}{\omega_{0}a}\frac{1+T^{\ast}_{c}/T}{1+T^{\ast}_{a}/T}T.
\label{eq:cond5}\end{eqnarray}
Clearly the resistivity crosses over from $T^{2}$ to $T$ when $T$ rises above
$T^{\ast}_{a}$, while $\cot\theta_{H}$ varies throughout $T^{n}$ with
$n=0,1,2,3$ but only as $T^{2}$ when
$T\ll {\rm min}\left(T^{\ast}_{a},T^{\ast}_{b}\right)$. The linear $\rho$ and 
the quadratic $\cot\theta_{H}$  can only be attained asymptotically in the 
mutually excluding temperature ranges above and below $T^{\ast}_{a}$.
Furthermore, $R_{H}$ and $\sigma_{H}$ will never exhibit the observed, simple
power-law behaviors which are $T^{-1}$ and $T^{-3}$ respectively. The
thermopower is likely to come out linear in $T$ and with the correct sign of 
the slope($e<0$), however, the positive intercept observed in most experiments 
is missing. 

These general features may be illustrated in the concrete example of
YBa$_{2}$Cu$_{3}$O$_{7}$, when using the bandstructure of Ref.~\cite{andersen}
in a numerical parametrization of the Fermi surface. To fix the relative
lengths of flat parts and corners, we rely on the fit to ac Hall data in
Ref.~\cite{zheleznyak} fixing $b$ to 0.71. Since the ac data show more structure,
this choice seems more reasonable than making an independent best fit of
the resulting temperature dependences. With $b=0.71$, we find $a=0.88$ 
(consistent with Ref.~\cite{zheleznyak}) and $c=0.87$, implying that
$T^{\ast}_{a}\approx T^{\ast}_{c}\approx 0.15 \eta W_{H}$ and
$T^{\ast}_{b}\approx 0.65 \eta W_{H}$.
\begin{figure}[t]
\epsfxsize=8.5cm
\epsfysize=7.5cm
\centerline{\epsfbox{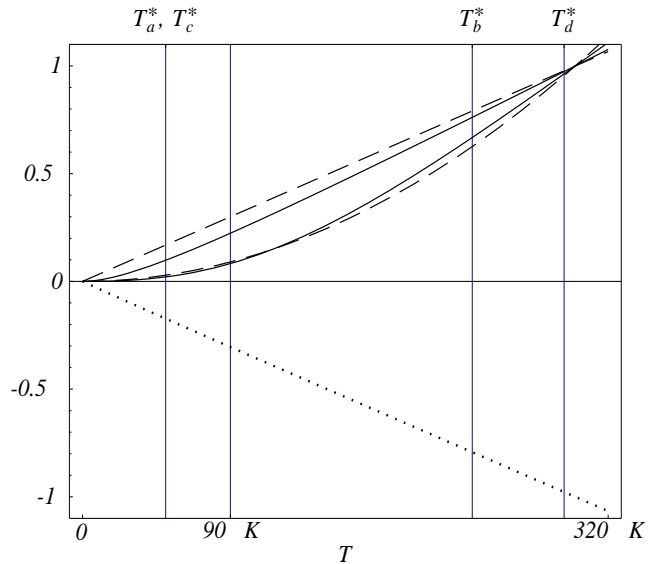}}
\caption[Figure 2]{\protect
$\rho(T)$ (upper full line),$\cot\theta_{H}(T)$ (lower full line) and $S(T)$ 
(dotted line) normalized to their values at room temperature, calculated with an
anisotropic scattering rate and using $\eta=4.5$ and $W_{H}=82.5$ K. Dashed 
lines show respectively $T/300K$ (upper) and $T^{2}/(300K)^{2}$ (lower) for
comparison.}
\label{fig2}
\end{figure}
Fig.~\ref{fig2} shows the resulting temperature dependences of $\rho$,
$\cot\theta_{H}$ and $S$, scaled to their values at room temperature an
using $\eta=4.5$ and $W_{H}=82.5$ K consistent with the fit of the additive
two-$\tau$ model to ac Hall data made in Ref.~\cite{zheleznyak}. 

Although the plotted curves for the resistivity and the inverse Hall angle are  
close to the wanted dependences of Eq.($\!\!$~\ref{eq:relaxt}) for the
parameters given and the temperatures considered, the explicit expressions for
$\rho(T)$ and $\cot\theta_{H}(T)$ from Eqs.($\!\!$~\ref{eq:cond4}),
($\!\!$~\ref{eq:hcond2}) show that, strictly speaking, these resemblances can
only occur in mutually excluding temperature ranges. Due to the proximity of
$a$ and $c$, the thermopower is largely linear in $T$ but with zero intercept.
With the neglect of the above-mentioned terms containing $\partial_{k\|}\tau$,
the calculated magnetoresistivity
$\Delta\rho/\rho=-\Delta\sigma/\sigma-\theta_{H}^{2}$ introduces an additional
temperature scale $T^{\ast}_{d}$ which we find to be $0.80 \eta W_{H}$ in the
example considered in Fig.~\ref{fig2}. It varies as $T^{-4}$ for
$T\ll T^{\ast}_{a,b,c,d}$, again in approximate agreement with the
experiment~\cite{harris}, but it crosses over to $T^{-2}$ as $T$ rises above
$T^{\ast}_{a,b,c,d}$.

In Refs.~\cite{carrington,kendziora} the additive two-$\tau$ approach was used 
together with Ong's formula for the Hall conductivity~\cite{ong,harris} as
being equal to the area traced out by the mean free path vector as one
circulates the Fermi surface. These areas were estimated to be proportional to
$l_{c}l_{f}$ whereby $\sigma\propto l_{f}$ ensures that
$\tan\theta_{H}\propto l_{c}$, in agreement with Eq.($\!\!$~\ref{eq:relaxt}).
We stress, however, that the exact area is given by the integral of $l^{2}$
which, in accordance with Eq. ($\!\!$~\ref{eq:hcond}), yields a sum of squares
of $l_{c}$ and $l_{f}$ rather than the product of the two.

A concrete realization of the hot-spot model is the nearly antiferromagnetic 
Fermi liquid (NAFL)~\cite{pines}(cf. also Ref.~\cite{andong} for points much
alike those being made here) in which the quasiparticle relaxation rate is
determined by the coupling to antiferromagnetic spin fluctuations. In
Ref.~\cite{pines} the Hall conductivity was found to contain not only the
$T^{-4}$ and $T^{-2}$ terms important, respectively, at low and high
temperatures, but also a $T^{-3}$ term which dominates in some intermediate
range of temperatures. This crucial $T^{-3}$ term appears to be a special
feature of the NAFL model, but is not intrinsic to the two-$\tau$ approach
which is based on the $T$-independent $a$ and $b$. 

It was pointed out by Hlubina and Rice~\cite{hlubina} that for microscopic
models featuring hot spots, the standard result for the resistivity consistent
with Eq.($\!\!$~\ref{eq:cond4}) can be improved by using a
{\it short-circuiting} variational solution of the Boltzmann equation much like
${\bf E}\cdot{\bf v}/(e^{|\Delta_{k_{\|}}|}+1)$, where $\Delta_{k_{\|}}$ is 
minimal on the cold corners, and vanishes all together when $T$ increases above 
$\eta W_{H}$ where the two rates become equal. This distribution function favors
the cold corners, enhances $T_{a}^{\ast}$ and extends the region of quadratic
resistivity to even higher temperatures, thereby further reducing the 
potentiality  of the hot spot model. It is worthwhile mentioning that the
recent results of Boebinger {\it et al}.~\cite{boebinger} indicate that the 
linear resistivity and quadratic $\cot\theta_{H}$ observed at optimal doping
indeed persist down to very low temperatures when suppressing superconductivity
with a large pulsed magnetic field.  

One might note that this short-circuiting of the Fermi surface bears a certain
resemblance to the behavior one would expect upon entering the underdoped region,
where a $d$-wave pseudogap has been shown to evolve~\cite{ding}. Thus, with a
$d_{x^{2}-y^{2}}$-wave symmetric gap in the quasiparticle spectrum
$E_{k}=\sqrt{\xi_{k}^{2}+\Delta_{k}^{2}}$, the conductivity will have enhanced
weights on the nodal regions, which is again restricting the linear resistivity
to higher temperatures. In the underdoped materials, a restriction of the
linear resistivity to higher temperature is indeed observed, but rather as the 
top of an s-shaped curve, than as a tangent to the low temperature parabola found
here~\cite{batlogg}.

As for the vanishing intercept in the thermopower, one possible remedy of the 
above might be to use an energy dependent relaxation rate causing an additional
term in $\beta$ through the expansion of $\tau(\varepsilon)$. This remains to be 
resolved, though it has been suggested in Ref.~\cite{cooper} that spin
fluctuations may accommodate just the right energy dependence of $\tau_{c}$ for 
the corners to yield the missing constant term.

To summarize, we have reinvestigated the hot-spot or additive two-$\tau$ model,
with an emphasis on temperature dependences of galvanomagnetic and thermoelectric
coefficients. In spite of a somewhat deceiving resemblance with the existing data,
we find systematic deviations of these dependences from the simple power-laws
which have been found as approximately universal features of the normal state
transport in optimally doped cuprates. As an immediate consequence of the explicit
formulae of Eqs.($\!\!$~\ref{eq:cond4}-$\!\!$~\ref{eq:cond5}), we note
the unavoidable appearance of additional temperature scales ($T^{\ast}_{a,b,c,d}$),
which to some extent obscures the original intention underlying the
two-relaxation-time ansatz. The dependences given by
Eqs.($\!\!$~\ref{eq:cond4}-$\!\!$~\ref{eq:cond5}) should make it possible to
contrast the experimental data against a generic additive two-$\tau$ model and 
to assess a general validity of such a phenomenological description. 

The authors are grateful to V. M. Yakovenko, H. D. Drew, H. Smith and A. Luther
for valuable discussions.

\end{document}